\documentclass[preprint]{aastex}

\newcommand{\etal}{{\frenchspacing\it et al. }}

\newcommand{\lsim}{\hbox{ \rlap{\raise 0.425ex\hbox{$<$}}\lower 0.65ex\hbox{$\sim$} }}
\newcommand{\gsim}{\hbox{ \rlap{\raise 0.425ex\hbox{$>$}}\lower 0.65ex\hbox{$\sim$} }}

\shorttitle{DPOSS Photometry}
\shortauthors{Gal \etal}

\begin{document}

\title{The Digitized Second Palomar Observatory Sky Survey (DPOSS) II: \\
 Photometric Calibration}

\author{R.\ R. Gal\altaffilmark{1}, R.\ R. de Carvalho\altaffilmark{2,4}, S.\ C. Odewahn\altaffilmark{3}, S.\ G. Djorgovski, A. Mahabal, R.\ J. Brunner,  P.\ A.\ A. Lopes\altaffilmark{2}}
\affil{Palomar Observatory, Caltech, MC105-24, Pasadena, CA 91125}
\altaffiltext{1}{Johns Hopkins University, Center for Astrophysical Sciences, 3701 San Martin Dr., Baltimore, MD 21218 \\ 
{\indent Email: rrg@pha.jhu.edu} }
\altaffiltext{2}{Observat\'orio Nacional, Rua Gal. Jos\'e Cristino,
77 - 20921-400, Rio de Janeiro, RJ, Brazil}
\altaffiltext{3}{Arizona State University, Dept. of Physics \& Astronomy,
Tempe, AZ 85287} 
\altaffiltext{4}{Currently at Osservatorio Astronomico di Brera, Via Brera 28, 20121 - Milano, Italy}
  
\begin{abstract}

We present the photometric calibration technique for the Digitized Second Palomar Observatory Sky Survey (DPOSS), used to create seamless catalogs of calibrated objects over large sky areas. After applying a correction for telescope vignetting, the extensive plate overlap regions are used to transform sets of plates onto a common instrumental photometric system. Photometric transformations to the Gunn $gri$ system for each plate, for stars and galaxies, are derived using these contiguous stitched areas and an extensive CCD imaging library obtained for this purpose. We discuss the resulting photometric accuracy, survey depth, and possible systematic errors.
\end{abstract}


\section{Introduction}

The answers to many important cosmological questions require large sky surveys, encompassing hundreds or thousands of square degrees. Large scale structure in the universe, the power spectrum at large angular scales, and the distribution of stars in our own galaxy all rely on homogeneous, moderately deep imaging with small (and well understood) systematic errors, as well as enormous sky coverage. Until very recently, such data was available only from photographic plates, which, with their large physical size, fill the focal plane of Schmidt telescopes. Unfortunately, these plates are notoriously difficult to calibrate, with large sensitivity variations among plates and even within a given plate. 
	Nevertheless, they provide the only current source of all-sky imaging data. Various projects, such as the Automated Plate Scanner (APS, \citealt{ode95}, hereafter OA95), have used digitized scans of the first Palomar Sky Survey, to generate large area catalogs. The more recent efforts by the Automatic Plate Measuring team (APM, \citealt{mad90}, hereafter MES ) and other groups have utilized the Southern Sky Survey plates or scans of the more recent Second Palomar Sky Survey for this purpose. Some of these projects have used only a fraction of the available data, with limited CCD calibration, which makes assessing large-scale photometric errors difficult, if not impossible. For instance, MES used a scheme to transform all of their plates onto a common photometric system, but only had CCD calibration data for about one third of their plates. In the Northern sky, the GSC-II \citep{mac98} and the USNO \citep{mon98} both provide catalogs (based on some of the same plate material), but they lack accurate photometric calibration and are not targeting faint objects. Only the recently begun Sloan Digital Sky Survey (SDSS, \citealt{yor00}), will provide improved data in the Northern sky, and only at galactic latitudes $|b|>40^{\circ}$.

	To address these issues, and provide the community with a large catalog of sources with accurate photometry, we have produced the Digitized Second Palomar Observatory Sky Survey \citep{djo99,djo02}. Prior papers have discussed the object detection and classification techniques, using the SKICAT software package \citep{wei95a,wei95c}, and a companion paper \citep{ode02} presents more recent Artificial Neural Network (ANN) and Decision Tree (DT) classifiers employed in this survey. \citet{gal00b} has presented the vast CCD imaging sample obtained at the Palomar 60-inch for the purpose of calibrating DPOSS. Here we discuss the techniques used to derive the plate photometric calibration, and generate a seamless catalog of objects over the high-galactic-latitude ($|b|>30^\circ$) Northern sky. We briefly review the salient details of the DPOSS and CCD object detection and photometry schemes in \S2. The use of plate overlap regions to transform sets of plates on a uniform instrumental magnitude system is described in \S3. The derivation of the photometric calibration using CCD data is discussed in \S4, including our final photometric errors. We conclude with a brief discussion of potential applications and future developments in \S5.
We note that this paper supersedes the earlier discussion of DPOSS calibration presented in \citet{wei95b}.

\section{DPOSS and CCD Data}

\subsection{DPOSS}
The POSS-II photographic survey \citep{rei91} covers the entire Northern sky ($\delta > -3^\circ$) with 897
overlapping fields (each $6.5^\circ$, with $5^\circ$ spacings), and,
unlike the old POSS-I, has no gaps in the coverage.  Approximately half
of the survey area is covered at least twice in each band, due to plate
overlaps.  Plates were taken at the Palomar 48-inch Oschin Schmidt telescope in three bands:
blue-green, IIIa-J + GG395, $\lambda_{\rm eff} \sim 480$ nm;
red, IIIa-F + RG610, $\lambda_{\rm eff} \sim 650$ nm; and
very near-IR, IV-N + RG9, $\lambda_{\rm eff} \sim 850$ nm. The bandpasses are illustrated in Figure 1.
Typical limiting magnitudes reached are $g_J\sim21.5$, $r_F\sim21.0$, and $i_N\sim20.3$, i.e., $\sim 1^m - 1.5^m$ deeper than the POSS-I.  The image
quality is improved relative to the POSS-I, and is comparable to the southern
photographic sky surveys.

The original survey plates are digitized at STScI, using modified PDS scanners \citep{las96}. The plates are scanned with
15-micron (1.0 arcsec) pixels, in rasters of 23,040 square, giving $\sim 1$
GB/plate, or $\sim 3$ TB of pixel data total for the entire digital survey
(DPOSS).  The preliminary astrometric solutions provided by GSC-II \citep{mac98} are good to r.m.s. $\sim 0.5''$, and are in the process of being improved substantially.

An extensive effort, centered at Caltech, and with sites in Italy (Osservatorio Astronomico di Capodimonte and Osservatorio Astronomico di Roma) and Brazil
(Observat\'orio Nacional), has resulted in the processing, calibration, and cataloging of nearly all scans at $|b|>10^{\circ}$, with the detection
of all objects down to the survey limit, and star/galaxy classifications
accurate to 90\% or better down to $\sim 1^m$ above the detection limit.
Object detection and photometry is performed by SKICAT, a novel software system developed for this purpose \citep{wei95b}.  It incorporates some standard astronomical image processing packages, commercial Sybase database, as well as a number of artificial intelligence (AI) and machine learning (ML) modules. We measure $\sim 60$ attributes per object on each plate in each filter. A subset of these are used for classification, as described in \citet{ode02}. 

\subsubsection{Vignetting Correction}

Large area detectors have always faced the problem of nonuniform 
illumination across the detector. The resulting effect, primarily due to
placing a square detector in a telescope with a round aperture, is called 
vignetting. The vignetting pattern must be removed if one is to use the data 
at larger radii from the plate center. A simple (but incorrect) approach is
to assume a circularly symmetric function and apply it
to the data. However, though largely radially symmetric, the pattern has 
finer structure, due to the bending of the plates, the filters used, etc.
Here we describe the procedure
we have used to remove the vignetting pattern from the DPOSS data.
This procedure is performed separately for the three filters used.

Basically, we stack and combine multiple DPOSS plate images to obtain a 
``master vignetting correction'' field, in a manner analogous to generating
a flat field for CCD data. Because this procedure was developed after most
plates were processed into catalogs, the vignetting correction image is used
 to derive corrections to the object magnitudes post-processing, as well as 
correct the actual pixel data. The following procedure was followed, using 
one hundred plates for each band:

\begin{itemize}
\item Bin each plate image $8\times8$, resulting in a $2880\times2880$ pixel image.
\item Normalize each image by the median of its central $720\times720$ region.
\item Stack the resulting images. This results in a $2800\times2800\times100$ array.
\item For each pixel in the array, discard the ten brightest and ten faintest members, and obtain the median of the remaining eighty. This produces a single $2880\times2880$ image, where each pixel is now a clipped median of the one hundred contributing images.
\item Normalize the resulting image such that the maximum in the central
$720\times720$ pixel region is 1.0. 
\end{itemize}

Thus, all corrections are relative to the plate center, which is the best exposed part of the plate. Contours of the master vignetting correction image in the $F$ and $J$ bands are shown in Figure 2; the $N$ map is similar. Further details of this procedure can be found in \citet{mah02}.

Nearly all plates at $|b|>10^{\circ}$ have been processed into catalogs. The final result of this processing is the Palomar-Norris Sky Catalog (PNSC), expected to contain $\sim 50\times10^6$ galaxies and $>2\times10^6$ stars.

\subsection{CCD Data}
Details of the acquisition, processing, and calibration of the CCD data used to calibrate DPOSS are provided in \citet{gal00b}.  Here we provide a brief overview of these data.

We have imaged nearly 900 independent CCD fields for DPOSS calibration. The imaging targets are selected from the list of northern Abell Clusters \citep{abe89}, with priority given to the richest clusters closest to the field centers. This strategy increase the number of objects for both star/galaxy separation and photometric calibration. For those plates with no Abell clusters, we image the plate center and two other pointings within the plate. All data are obtained at the Palomar Observatory 60-inch telescope with the CCD imaging cameras placed at Cassegrain focus.  The Gunn $gri$ filters \citep{thu76,wad79,sch83} are used, which are well matched to the DPOSS bandpasses (see Figure 1). Data are taken only on photometric nights with seeing better than $2''$. The mean seeing for our data is $\sim1.5''$, with limiting magnitudes of approximately $m_{lim}=21.5^m, 21.5^m, 21.2^m$ in $g$, $r$, and $i$, respectively ( $\sim0.5^m-1^m$ deeper than the plate detection limits). For every night that is deemed photometric, we observe a set of Gunn standards \citep{ken85}.

Data are processed in the usual way using the IRAF data reduction package \citep{tod86}. All frames are bias subtracted and flat fielded, using a combination of dome and twilight flats.  Additionally, a dark sky flat field correction is generated by median filtering all of the unregistered target frames in each filter on a given night. Finally, we apply a fringe correction to the $i$ images. The photometric standard stars observed on each night are photometered using the {\it apphot} package in IRAF, and the resulting collection of between five and twelve measured instrumental magnitudes used to determine the zero-point offset, airmass term, and color term in each filter.

Object detection on the target frames is performed using the FOCAS package \citep{jar81,val82}. The $g$, $r$, and $i$ frames are processed independently, using detection parameters of $2.5\sigma$ per pixel, a 25 pixel minimum area, and a sky value estimated individually for each image. Object classification is also performed by FOCAS, and the classifications visually inspected. Bright objects with incorrect classifications (usually due to saturated pixels) are corrected by hand. The photometric coefficients (zero point, color term, and airmass term) derived from the standard star observations are used to determine object magnitudes.

\section{Plate Overlap Analysis}

A great advantage of POSS-II over earlier surveys is the extensive overlap between neighboring plates. Adjacent plates overlap each other by $1.5^{\circ}$ along edges covering $6.5^{\circ}$, providing $\sim9 \Box ^{\circ}$ of duplicated area for each plate pair. This compares favorably to the APM survey, whose scans typically had $1^{\circ}$ overlaps \citep{mad90}; with the strong effect of vignetting at plate edges, the extra $0.5^{\circ}$ of overlap in DPOSS is significant. These areas each contain thousands of objects which can be used to derive the transformation of one plate's instrumental magnitude system to that of its neighbor. These transformations can be propagated across multiple boundaries to tie a contiguous set of plates onto a common photometric system.

In practice, this instrumental magnitude transformation must be mapped independently for stars and galaxies, especially for brighter magnitudes ($m_{\rm inst}<19$). Although the density-to-intensity (D-I) transformation has been linearized using the densitometry spots on each plate, stars and galaxies nevertheless display distinct photometric behaviors. This is because the brighter pixels in stellar images populate the highly nonlinear part of the D-I relation, where small errors in the polynomial fit (mostly due to the small number of points characterizing the ``shoulder'' of the curve) result in larger variations in the derived intensity. Pixels within galaxies occupy the more linear part of the D-I transformation, resulting in more stable and linear photometry. 

Additionally, propagating the photometric transformations across a large number of plates (as would be necessary to knit the whole sky) can result in the unacceptable accumulation of (possibly systematic) errors. This issue was addressed by \citet{gro86},  and their technique utilized by MES to estimate the extent of such errors in the APM. A similar approach was also taken by OA95 in an analysis of galaxy counts at the North Galactic Pole, but that study was limited by the small overlap regions and poorer quality of POSS-I. An early attempt to address the problem of plate inhomogeneities in POSS-II is discussed in \citet{pic91}. 

To avoid this issue entirely, we have elected to calibrate each plate using a sliding boxcar technique, described below. We transform each plate's eight surrounding neighbors onto the instrumental system of the center plate, match all the CCD data available for the resulting nine-plate area, and derive the photometric calibration for the center plate. The individually calibrated plates can then be quilted with no additional offsets because they are already on a calibrated, rather than instrumental, magnitude system. Unlike MES and OA95, we are allowed this luxury by the large quantity of available CCD calibration data. Figure 3 illustrates the processed DPOSS plates (squares), and the available CCD data (dots).

First, we determine which plate pairs have overlap regions larger than $\sim5 \Box ^{\circ}$, which removes pairs where only the heavily vignetted corners overlap. The overlap regions are matched, and the outermost 10\% of the matched area is clipped on all four sides, to avoid the most heavily vignetted regions. A spline is fit to the magnitude difference of the matched objects from the two plates, as a function of instrumental magnitude, $\Delta m = m_{\rm plate1}-m_{\rm plate2} = f(m_{\rm plate1})$. This fit is performed independently for stars and galaxies in each filter ($JFN$). At faint magnitudes ($m_{\rm inst}>19.0$) we use both stars and galaxies in the independent fits, to avoid biases due to possible differences in classification between plates. At these faint magnitudes, there are no bright pixels (which populate the highly nonlinear part of the D-I relation) in the stellar images. 

There are typically between 5,000 and 30,000 objects used to determine the transformation between a given plate pair. Although significant scatter is present, especially at faint magnitudes, the mean is nevertheless well determined. At the bright end ($m<16$), where the number of objects is small and classification problems exist, points for the spline fit are placed by hand. At intermediate magnitudes (typically $16<m<20$), a robust locus mapping algorithm is used. Finally, at the faint end ($m>20$), where incompleteness is a factor, the fit is fixed at a constant $\Delta m$. These objects are in any case too faint to be photometered and classified properly, and are therefore of limited scientific use from the current survey. Figure 4 shows a typical pair of plate overlaps, for galaxies (top) and stars (bottom) matched between plates 392 and 333. The resulting spline fits are also shown.

In contrast, MES defined a single coefficient, $T_{ij}$, to match the magnitudes between adjacent plates. While they state that the $rms$ of a linear fit to the magnitude differences (in their case, $(m_i-m_j)$ {\it vs.} $(m_i+m_j)/2$) is nearly as good as a higher order polynomial, we find that this is not always the case, and is certainly not true for stars (which they do not treat). We therefore have elected to utilize our procedure as described above, which is more general, but requires a great deal more interaction and inspection. This process also permits us to find individual plates which behave abnormally, either due to their intrinsic properties or processing failures. 

To actually transform a given plate's instrumental system to that of the reference plate for a given area, we propagate magnitude transformations through sets of up to five plate overlaps connecting the plate being transformed to the fiducial plate. This is illustrated in Figure 5, where arrows represent the transformation paths used to tie plates 392 and 452 to the center (fiducial) plate 391. A plate directly adjacent to the center plate can be tied to it by up to four paths (the green/solid arrows), while a plate diagonally away from the center can be tied by up to five paths (the red/dashed arrows). The five plate limit is imposed to avoid propagating errors across a large number of plate overlaps. Figure 6 shows the five photometric transformations derived between plates 331 and 391. The ordinate shows the fiducial plate magnitude $m_f$, while the abscissa shows the difference between the magnitude on the plate to be transformed, $m_p$, and $m_f$. Rather satisfyingly, all possible convolution paths provide similar transformations for a given plate, especially for galaxies; the $rms$ of the differences of each path from the mean reflect the scale of photometric errors due to plate pistoning.  These transformations are averaged for each plate to generate a spline $\Delta m_{pf}=m_p-m_f$ as a function of $m_f$, shown as the solid line in Figure 6. These plots are visually inspected for each plate pair, and instances of abnormal behavior, such as highly variable transformations, are flagged and treated by hand. This technique allows us to detect plate overlaps with unusual behaviors, and provides more stable magnitude matching than possible using a single overlap. The transformations for stars are highly nonlinear, and show much more variance than those for galaxies. This contributes to the large photometric errors in stellar photometric calibration discussed in \S4.2. Figure 7 demonstrates the removal of plate-to-plate pistoning using this technique. The left panel shows galaxies in an uncorrected instrumental magnitude range of $0.5^m$, from an area covering 15 plates. Individual plates are evident as square patches of varying galaxy density. The right panel shows galaxies from the same area, after the instrumental magnitudes have been homogenized; individual plates are no longer discernible. 

\section{Calibration to the Gunn $gri$ System}

Once a set of nine plates has been transformed to a common instrumental system (that of the center plate), we match our CCD catalogs to the plate catalogs for that area. Typically, we have between 15 and 50 CCD pointings, in the three filters ($gri$), for a given nine plate area. This provides between 8,000 and 30,000 matched galaxies and a similar number of stars from which to derive photometric calibration.

The CCD and plate data are matched, allowing for an overall offset, scale change, and rotation between the CCD-derived celestial coordinates and the plate celestial coordinates. After the mean CCD-plate coordinate offset is derived, a search radius of $3''$ is used, and the search process iterated until the RA,Dec offsets converge. Typical coordinate residuals are $0.4''$, which both confirms the stated astrometric accuracy of DPOSS, and implies that false matches are not a significant source of added noise. A spline is fit to the difference between the CCD calibrated magnitude $m_{\rm CCD}$ and the plate instrumental magnitude $m_{\rm inst}$, in a manner identical to the plate overlap analysis. Once again, this is done independently for stars and galaxies, relying on the plate-based classification (as this classification ultimately determines which calibration will be applied). Figure 8 shows a typical set of points and the resulting transformations between plate $F$ and CCD $r$ magnitudes, for both stars and galaxies, in a nine plate area. A linear fit is also shown; the nonlinear spline fits are clearly necessary. The derived transformations from $J\rightarrow g, F\rightarrow r, N\rightarrow i$ are then applied solely to the center field (391 in the example here).
 
This method differs sharply from that employed by MES, who obtained only magnitude zero points from their CCD data. Their frames, at $2'\times3'$ in size, cover only 6\% of the area of our CCD images, providing a factor of 17 fewer calibrating objects from detector area alone; considering our deeper exposures and more numerous CCD fields (nearly $900$ {\it vs.} $69$), we have a factor of approximately one-thousand more calibration objects. The nonlinearities in the transformations to calibrated magnitudes, if not treated, can lead to incorrect number count slopes and correlation functions. 

\subsection{Color Terms}

An examination of the filter curves in Figure 1 shows that the $F\rightarrow r$ and $N\rightarrow i$ transformations are likely to be free of color terms, as the plate and CCD filter systems are well matched. However, the $J$ filter is significantly broader and bluer than the corresponding CCD $g$ filter, and we may expect a color correction to be necessary. In the left panels of Figure 9, we show the difference between calibrated plate magnitude and CCD magnitude, $g_0(J)-g_{\rm CCD}$ as a function of $g-r$ color, for both galaxies and stars. A strong color term is evident, especially for the stars. The galaxies do not span a significant range in $g-r$, making the color term impossible to discern in the noise. Nevertheless, this color correction is extremely important for astrophysical purposes, such as photometric redshift estimation, galaxy typing, and cluster detection.

To estimate the color correction, we calculated the median $g_0(J)-g_{\rm CCD}$ in bins of $\Delta(g-r)=0.1$, shown as the filled squares in Figure 9. We fit a linear relation to these points, resulting in a corrected magnitude
\begin{equation}
g_{\rm corr}(J)=g_0(J)-0.373\times[g_0(J)-r_0(F)]+0.177
\end{equation}
We apply this correction to both stars and galaxies, with the results shown in the right panels of Figure 9. The correction both removes the slope and reduces the scatter in the relations, as expected. 

Inspection of the corresponding data for $r(F)$ and $i(N)$ shows that no color terms are needed for those filters, as expected from the well-matched filters for the plates and CCDs.

\section{Photometric Accuracy}

Photometric errors in a plate-based survey consist of random errors within plates, zero-point variations within a plate, and plate-to-plate zero point offsets, also called pistoning. Random errors are due simply to Poisson noise. Zero-point variations within individual plates are due both to telescope vignetting and inhomogeneous emulsion sensitivity; these are discussed in detail in \citet{mah02}. Plate pistoning is due to both varying emulsions between plate batches, and differences in observing conditions, such as moonlight and seeing.  Using both our extensive CCD calibration set, and the large plate overlap regions, we can measure the magnitude of these effects.

\subsection{Comparison with CCD Data}

First, we compare the calibrated plate magnitudes from a large number of plates with the corresponding CCD data. The errors measured in this procedure will include random plate photometric errors, plate-to-plate pistoning, as well as random errors from the CCD photometry. Using the internal tests of the CCD data presented in \citet{gal00b}, the tests in this section, and the plate overlap tests from the next section, we can decompose our measured errors into the constituent sources.

We plot the plate {\it vs.}~CCD magnitudes in Figure 10.  The $1\sigma$ errors for galaxies are $0.21^m,0.19^m,0.32^m$ at $m=19.0$ in the $g,r$ and $i$ filters, respectively. Table 1 provides the average photometric errors for galaxies and stars as a function of magnitude for the three bandpasses; these are plotted in Figure 11.

\begin{deluxetable}{ccccccccc}
\tablecolumns{9} 
\tablewidth{0pc} 
\tablecaption{Total Photometric Errors \label{tbl-1}}
\tablehead{
\colhead{}    & \multicolumn{2}{c}{$g$}  & & \multicolumn{2}{c}{$r$}  & &\multicolumn{2}{c}{$i$} \\
\cline{2-3}  \cline{5-6}  \cline{8-9} \\[-10pt]
\colhead{Mag} & \colhead{gal} & \colhead{star} & & \colhead{gal} & \colhead{star} & & \colhead{gal} & \colhead{star} } 
\startdata
17.0 & 0.25 & 0.22 & & 0.13 & 0.15 & & 0.17 & 0.22 \\
18.0 & 0.17 & 0.24 & & 0.16 & 0.17 & & 0.24 & 0.25 \\
18.5 & 0.19 & 0.25 & & 0.17 & 0.22 & & 0.29 & 0.32 \\
19.0 & 0.21 & 0.24 & & 0.19 & 0.21 & & 0.32 & 0.31 \\
19.5 & 0.25 & 0.26 & & 0.24 & 0.24 & & 0.36 & 0.34 \\
20.0 & 0.29 & 0.29 & & 0.29 & 0.27 & & 0.44 & 0.35 \\
20.5 & 0.37 & 0.32 & & 0.34 & 0.33 & & 0.42 & 0.39 \\
\enddata
\end{deluxetable}

This scatter is an accurate reflection of our accumulated errors. Because the CCD data cover hundreds of plates, and come from hundreds of nights, all sources of error are present in this calibration. This includes errors in the zero points of the CCD sequences, star/galaxy separation errors, inhomogeneities within individual plates, and errors in transforming all the plates to the fiducial instrumental system. Although the scatter is larger than those typically reported in the literature, it is the true combination of all possible errors. There are no additional effects to model or theoretical considerations to account for. These errors can therefore be used in any future work with no modification.

\subsection {Plate Overlap Tests}

The second test of photometric accuracy is a comparison of calibrated magnitudes from the overlap regions of two or more plates. This test is especially useful for examining photometric pistoning between the individually calibrated plates. A typical comparison in the $r$ band is shown in Figure 12. The results of this test show that for well-calibrated plates (those with over one thousand calibration galaxies), the mean plate-to-plate zero point offset is zero, with a $1\sigma$ scatter of $0.07^m$ in both the $g$ and $r$ bands, with only a slight magnitude dependence, as shown in Table 2. A few plate overlaps show much larger offsets (up to $\sim0.2^m$); these indicate photometric errors due to sensitivity gradients across some individual plates. Such plates constitute $\sim5\%$ of the overall sample. Calibration in the $i$ band is significantly worse, with much larger sensitivity variations both within and between plates, and many fewer calibrating objects.

\begin{deluxetable}{cccc}
\tablecolumns{4} 
\tablewidth{0pc} 
\tablecaption{Plate-to-Plate Offsets \label{tbl-2}}
\tablehead{
\colhead{}    &  \multicolumn{3}{c}{$1\sigma$ Scatter} \\[1pt]
\cline{2-4} \\[-8pt]
\colhead{Mag} & \colhead{$g$} & \colhead{$r$} & \colhead{$i$} }
\startdata
18.0 & 0.058 & 0.066 & 0.14 \\
18.4 & 0.059 & 0.070 & 0.14 \\
18.8 & 0.065 & 0.070 & 0.16 \\
19.2 & 0.071 & 0.073 & 0.13 \\
19.6 & 0.075 & 0.076 & 0.12 \\
20.0 & 0.080 & 0.078 & 0.13 \\
20.4 & 0.078 & 0.075 & 0.23 \\
\enddata
\end{deluxetable}

The distributions of plate-to-plate zero point offsets are well approximated by Gaussians with mean of 0 and standard deviations from Table 2, as shown in Figure 13. Therefore, users of DPOSS data may generate simulations including plate pistoning using the values provided in Table 2. 
 
Additionally, we performed the same test as with the CCD data, using the plate overlap regions. We examined magnitude differences for all pairs of matched objects for plate pairs, in the same magnitude bins as with the plate {\it vs.}~CCD test. We then assumed that the measured photometric uncertainties are due to equal contributions from both plates, and therefore divided the results by $\sqrt{2}$. The photometric uncertainties associated with individual objects, estimated in this way, are in complete agreement (to better than $10\%$, or $0.01^m$) with the tests comparing the plate and CCD data. This is expected, as the errors in the plate-to-CCD comparison are dominated by the poorer plate photometry. We again conclude that the photometric uncertainties stated in Table 1 are robust.

\section{Conclusions}

We have presented the photometric calibration technique employed in DPOSS. Due to the large plate overlaps in the POSS-II, and vast quantities of CCD calibration, we are able to perform accurate calibration over the entire high galactic latitude Northern sky. Our technique avoids significant systematic plate-to-plate zero-point offsets, resulting in stable photometry over extremely large angles. We provide the reader with robust photometric error measurements which can be used during scientific analysis of the data. Our results are consistent with the plate-to-plate zero point errors found by \citet{wei95b}, although their sample was much smaller.

The resulting photometric catalogs, along with star/galaxy classifications (Odewahn \etal 2002) can be used for a large variety of astronomical projects which require large sky coverage. These include the search for $z>4$ quasars \citep{ken95}, the Northern Sky Optical Cluster Survey \citep{gal00a,gal02}, searches for small groups \citep{iov02}, galaxy number counts, correlation functions, studies of galactic structure, and many more.

The catalog data from DPOSS is now publicly available via the web at {\url http://dposs.caltech.edu}. Additionally, the processed CCD imaging data used for calibration is available via the San Diego Supercomputing Center's Storage Resource Broker (SRB); directions for downloading this data are also at the aforementioned URL.

\acknowledgments

We thank the Norris Foundation for their generous support of the DPOSS project. RRG was supported in part by an NSF Fellowship, NASA GSRP NGT5-50215, and a Kingsley Fellowship. This work would have been impossible without the POSS-II photographic team and the STScI digitization team. We also thank the Palomar TAC and Directors for generous time allocations for the DPOSS calibration effort. Numerous past and present Caltech undergraduates (V. Desai, V. Hradecky, J. Meltzer, B. Stalder, J. Hagan, R. Stob, J. Kollmeier, B. Granett) assisted in the taking of the data utilized in this paper. This work was made possible in part through the NPACI sponsored Digital
Sky project and a generous equipment grant from SUN Microsystems. Access to the POSS-II image data stored on the HPSS, located at the California Institute of Technology, was provided by the Center for Advanced Computing Research. The SDSC has generously allowed placement of our CCD data into their SRB as part of the National Virtual Observatory (NVO) initiative.

\clearpage

\begin{figure}
\plotone{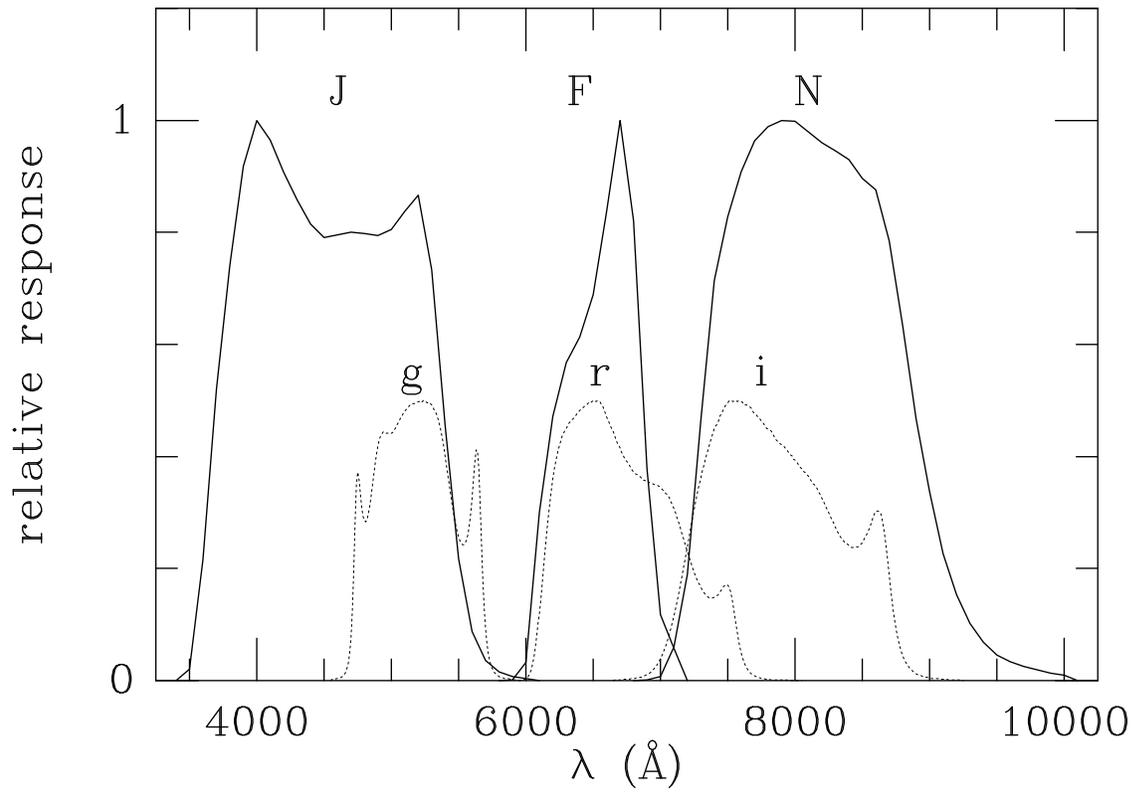}
\vskip -1.4truein
\caption{The POSS-II $JFN$ (solid lines) and Gunn $gri$ (dashed lines) bandpasses.}
\end{figure}

\begin{figure}
\plotone{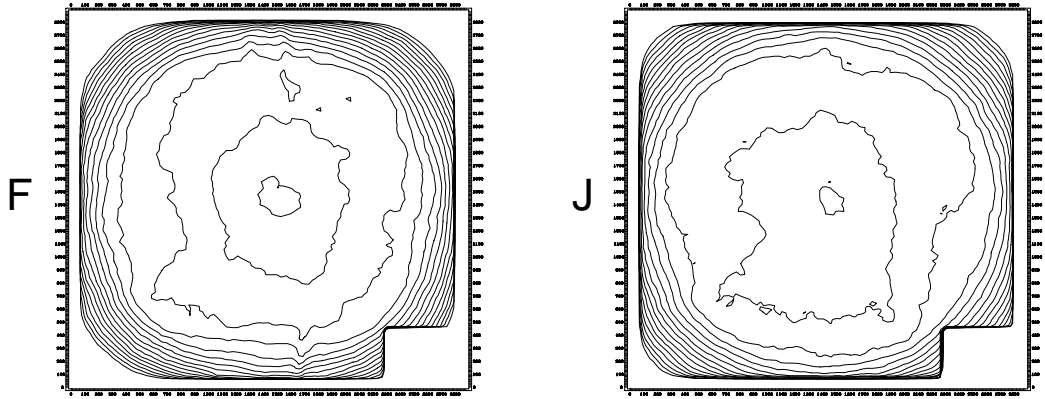}
\caption{Contour maps of the vignetting correction function for the $F$ and $J$ bands.}
\end{figure}

\begin{figure}
\plotone{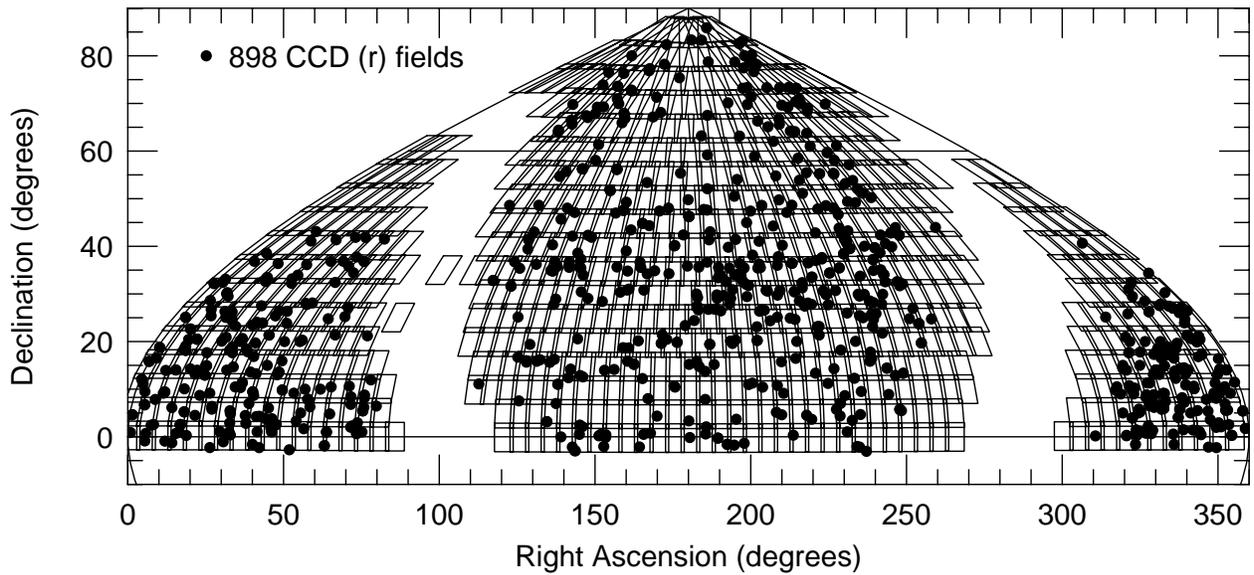}
\caption{Processed DPOSS plates (squares) and available CCD calibration pointings (dots).}
\end{figure}

\begin{figure}
\plotone{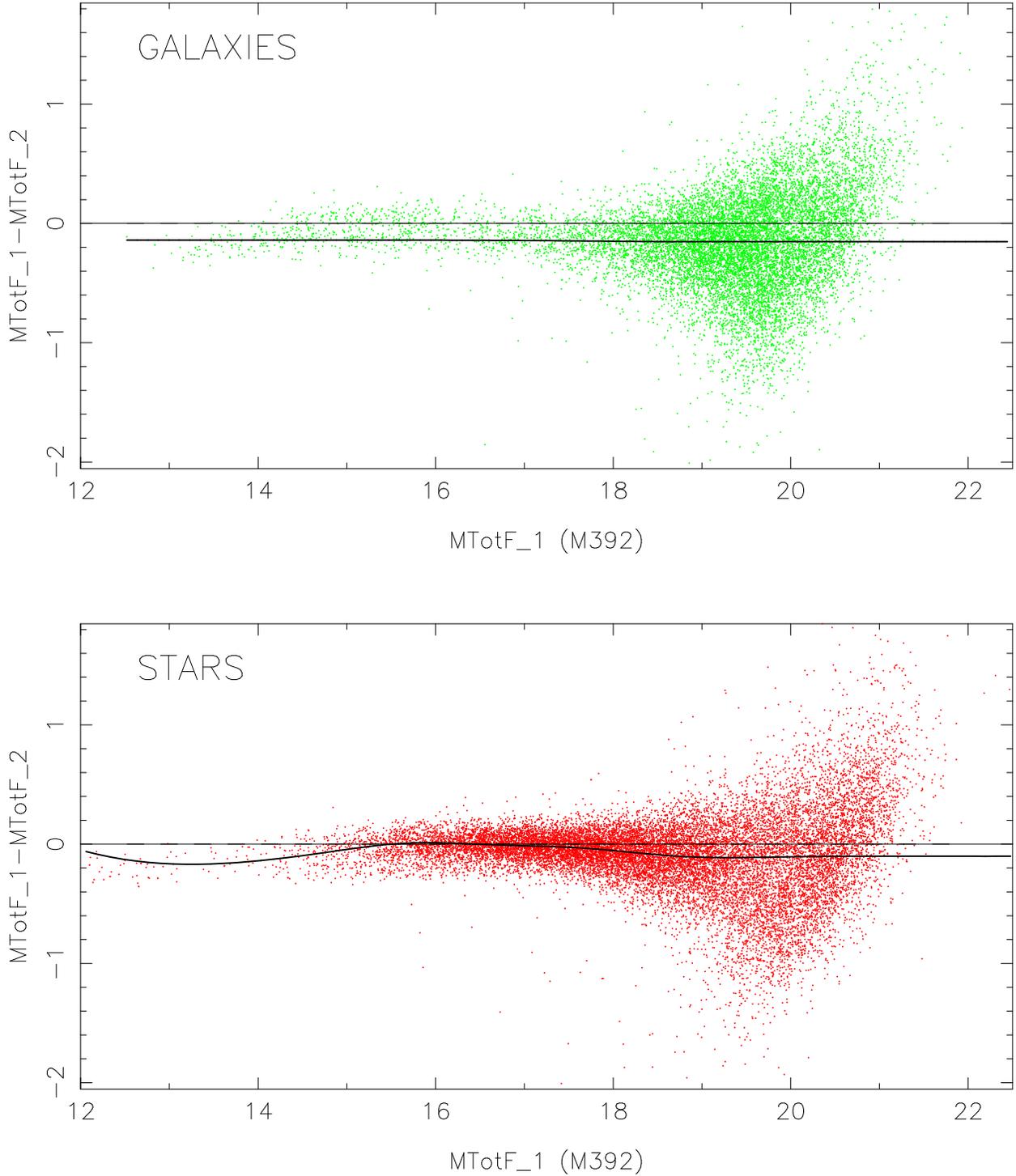}
\caption{A typical plate overlap analysis, for F plates in fields 333 and 392. We show the magnitude difference between plates for galaxies (top) and stars (bottom), along with the spline fits to unify the photometric systems. Brighter than $m=16$, the galaxy sample is heavily contaminated by misclassified stars, so the spline is constrained manually.}
\end{figure}

\begin{figure}
\plotone{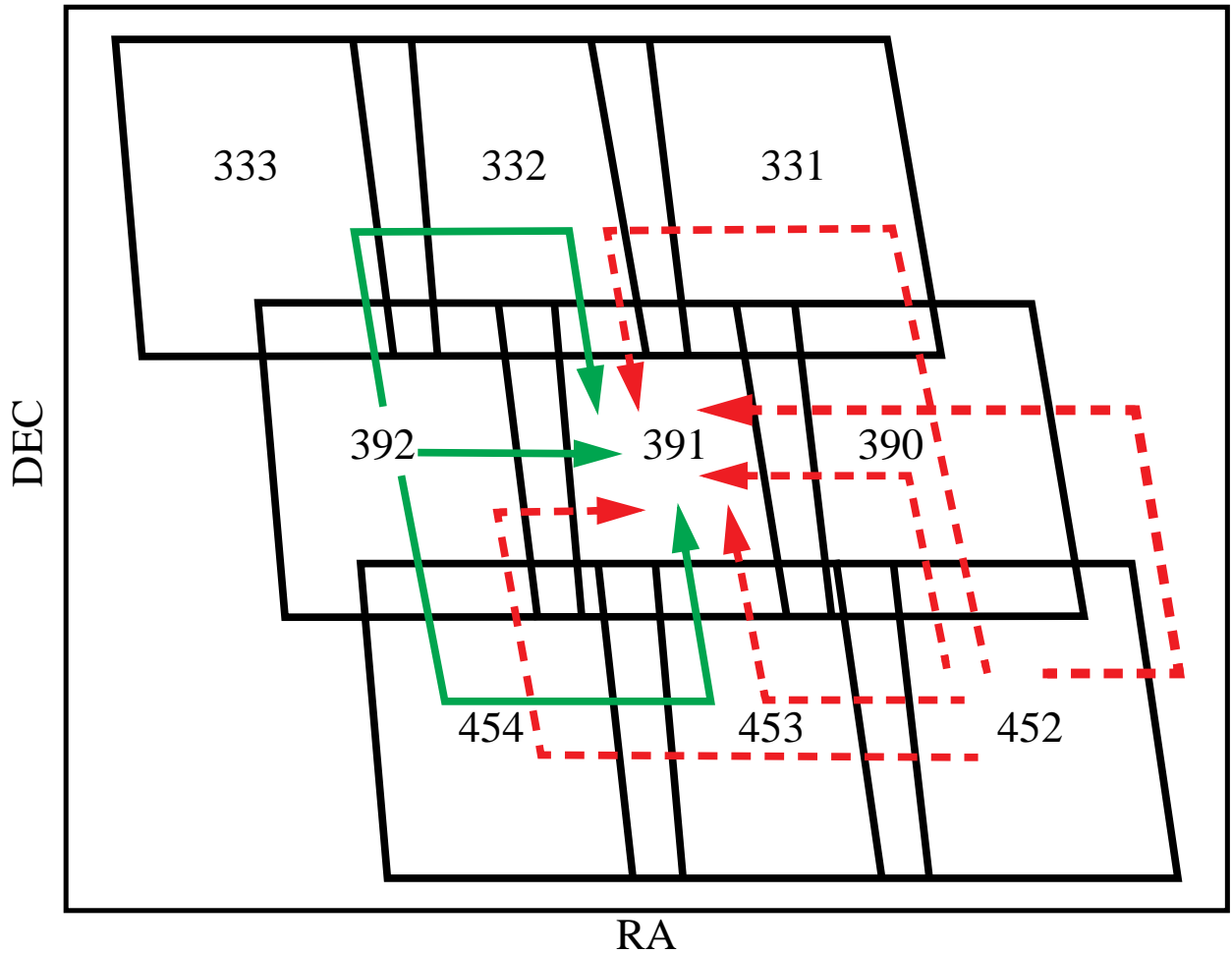}
\caption{Plate overlaps used to derive mean instrumental photometric transformations. Arrows represent the transformation paths used to tie fields 392 and 452 to the center (fiducial) plate 391. A plate directly adjacent to the center plate can be tied to it by up to three paths (the green/solid arrows), while a plate diagonally away from the center can be tied by four paths (the red/dashed arrows).}
\end{figure}

\begin{figure}
\plotone{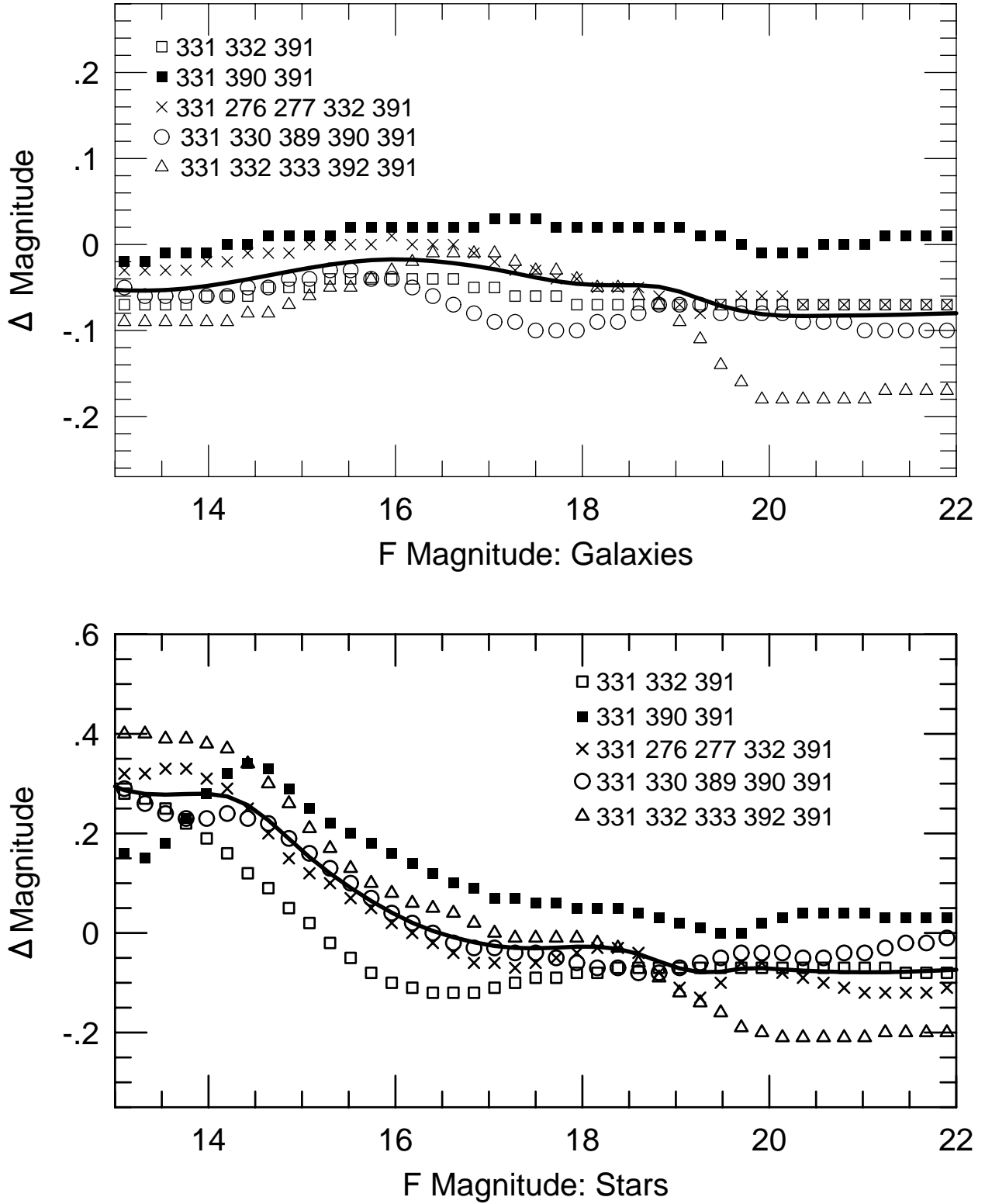}
\caption{The results of the overlap analysis illustrated in Figure 4. The ordinate shows the fiducial plate magnitude $m_f$, while the abscissa shows the difference between the magnitude on the plate to be transformed $m_p$ and $m_f$. Rather satisfyingly, all possible transformation paths provide similar transformations for a given plate.}
\end{figure}

\begin{figure}
%\plotone{Gal.photo.fig7.epsi}
\caption{Galaxies from a set of 15 plates before and after removal of plate pistoning. The left panel shows galaxies in an uncorrected instrumental magnitude range of $0.5^m$, from an area covering 15 plates. Individual plates are evident as square patches of varying galaxy density. The right panel shows galaxies from the same area, after the instrumental magnitudes have been homogenized; individual plates are no longer discernible. }
\end{figure}

\begin{figure}
\plotone{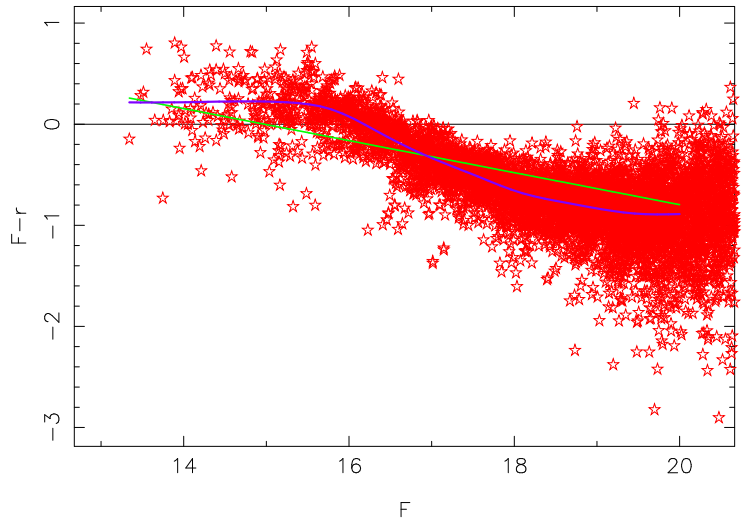}
\vskip 0.35truein
\caption{The transformation from DPOSS $F$ magnitudes to Gunn $r$ magnitudes derived for a set of 9 plates, for galaxies (top) and stars (bottom). A linear fit, much less a single zero-point offset, is clearly insufficient.}
\end{figure}

\begin{figure}
\plotone{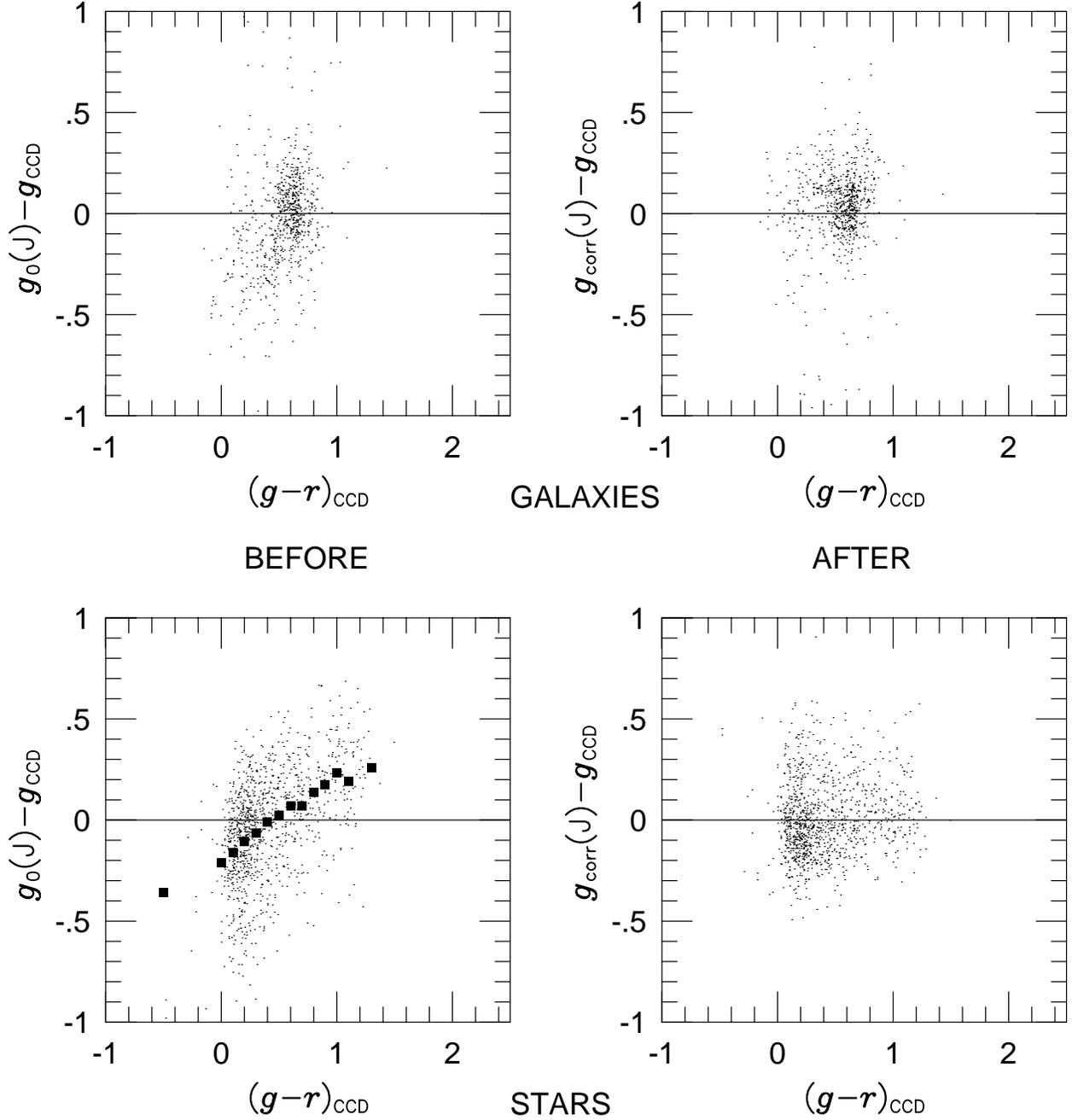}
\caption{The difference between CCD and calibrated plate magnitudes, $g(J)-g_{CCD}$, as a function of $g-r$ color, for galaxies (top) and stars (bottom), before and after a color correction has been applied. The filled squares show the median points used to derive the color correction.}
\end{figure}

\begin{figure}
\plotone{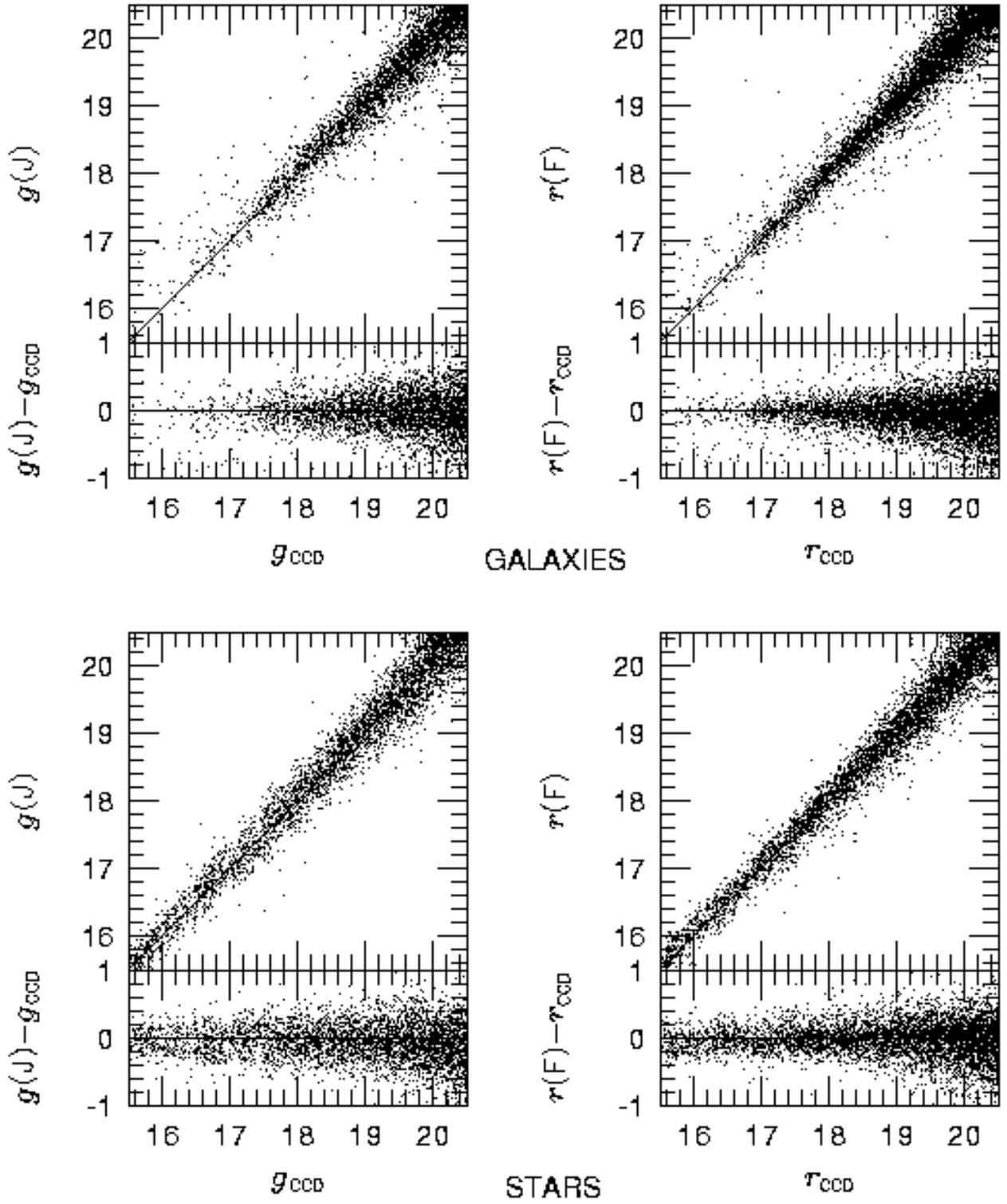}
\caption{Comparison of the DPOSS $g$ and $r$ magnitudes derived from the transformation shown in Figure 5, {\it vs.}~the corresponding CCD magnitudes, for galaxies (top) and stars (bottom).}
\end{figure}

\begin{figure}
\plotone{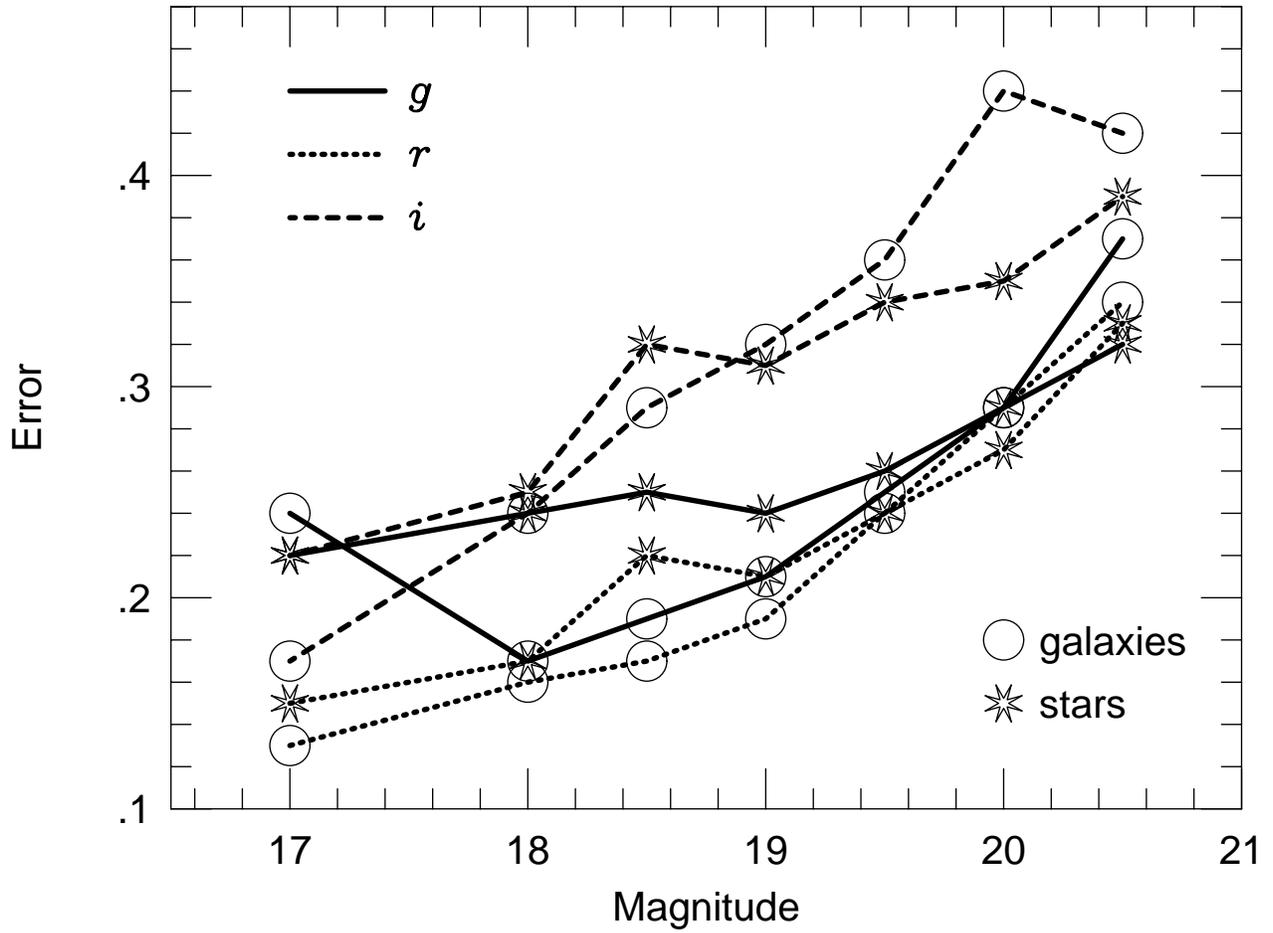}
\caption{Total photometric errors as a function of magnitude, for stars (star symbols) and galaxies (open circles), in $g$ (solid lines), $r$ (dotted lines), and $i$ (dashed lines).}
\end{figure}

\begin{figure}
\plotone{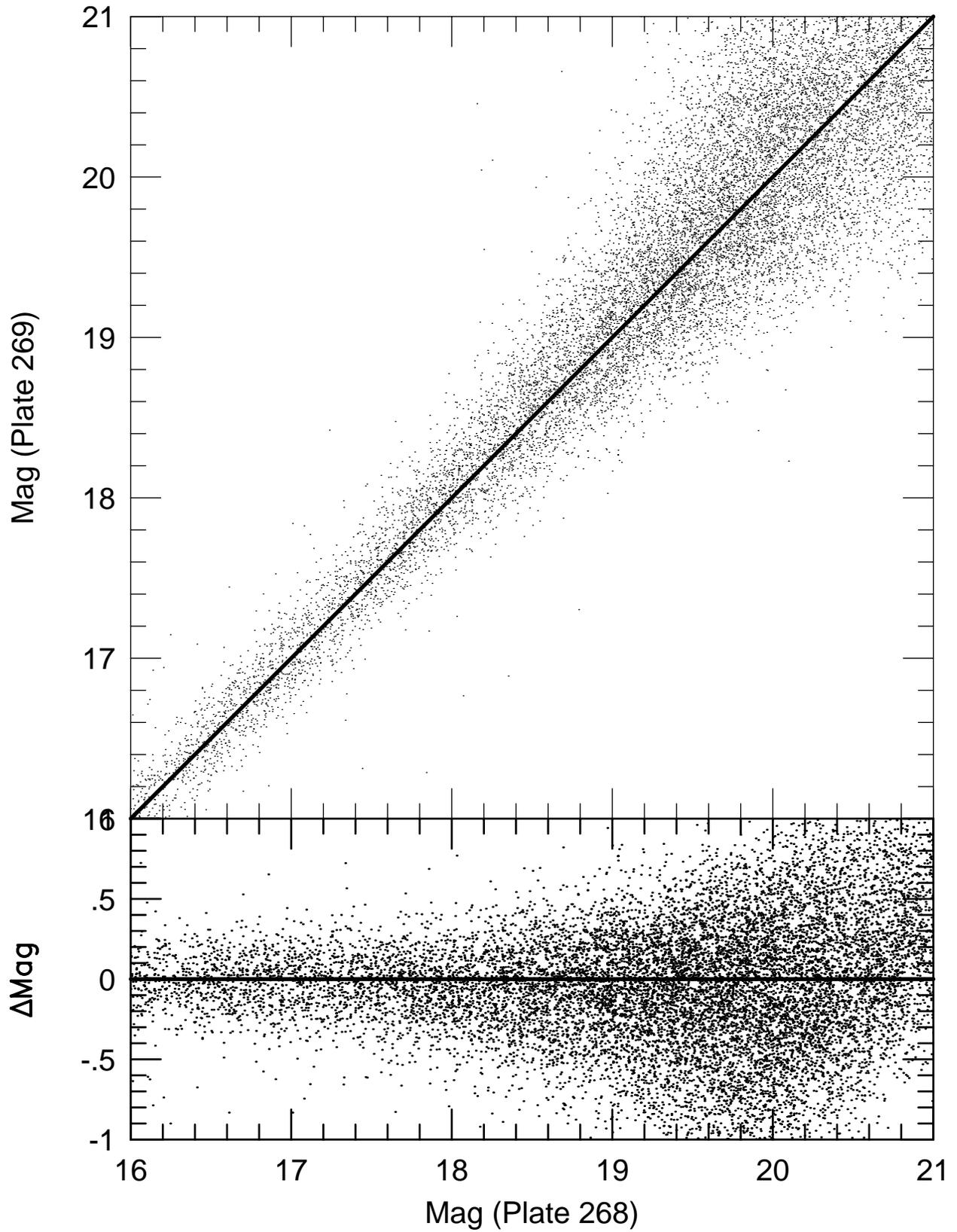}
\caption{Comparison of $r$ calibrated magnitudes from the overlap regions of plates 268 and 269.}
\end{figure}

\begin{figure}
\plotone{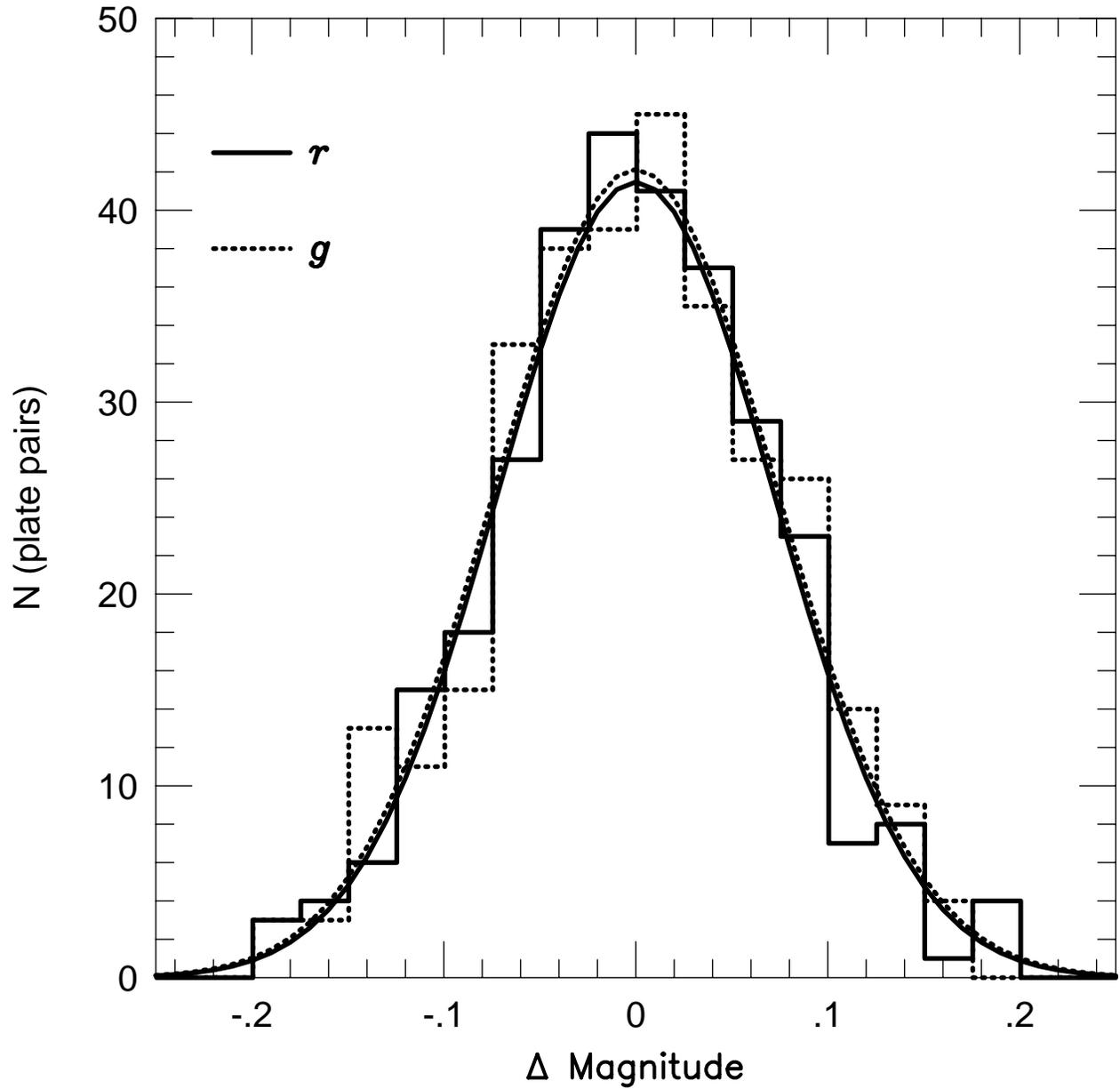}
\caption{The distribution of plate-to-plate zero point offsets at $m=19.4$ in the $r$ band (solid line) and $g$ band (dotted line). Gaussians with zero mean and $\sigma_{g,r}$ taken from Table 2 provide a good representation of the data.}
\end{figure}

\end{document}